\documentclass{amsart}
\usepackage{graphicx}

\newcommand{\ZZ}{\mathbf{Z}}
\newcommand{\ep}{\varepsilon}
\newcommand{\ceil}[1]{\left\lceil{#1}\right\rceil}
\newcommand{\floor}[1]{\left\lfloor{#1}\right\rfloor}

\DeclareMathOperator{\GF}{GF}

\newtheorem{thm}{Theorem}
\newtheorem{prop}[thm]{Proposition}

\theoremstyle{definition}
\newtheorem{example}[thm]{Example}
\theoremstyle{remark}

\begin{document}

\title[Multipoint Kronecker substitution]{Faster polynomial multiplication via multipoint Kronecker substitution}
\author{David Harvey}
\begin{abstract}
We give several new algorithms for dense polynomial multiplication based on the Kronecker substitution method. For moderately sized input polynomials, the new algorithms improve on the performance of the standard Kronecker substitution by a sizeable constant, both in theory and in empirical tests.
\end{abstract}

\maketitle

\section{Introduction}
\label{sec:intro}

The \emph{Kronecker substitution method} is an algorithm for computing the product of two polynomials. The basic idea was originally suggested by Kronecker \cite{kronecker} as a means of reducing problems concerning multivariate polynomials to univariate polynomials; later Sch\"onhage \cite{schonhage} suggested a similar idea to reduce multiplication in $\ZZ[x]$ to multiplication in $\ZZ$. The technique is widely used in practice: for example, the Magma computer algebra system uses Kronecker substitution to multiply polynomials in $\ZZ[x]$ in some cases \cite{magma}, and Victor Shoup's NTL library \cite{ntl} uses Kronecker substitution to reduce multiplication in $\GF(p^n)[x]$ to multiplication in $\GF(p)[x]$.

The reduction from $\ZZ[x]$ to $\ZZ$ is best explained by an example. To compute the product
 \[ h(x) = (621x^3 + 887x^2 + 610x + 274) \times (790x^3 + 424x^2 + 298x + 553), \]
one evaluates the polynomials at $x = 10^7$, and then computes the \emph{integer} product
\begin{align*}
 h(10^7) & = 621000088700006100000274 \times 790000042400002980000553 \\
   & = 490590096403410430461082839078846704189820151522.
\end{align*}
The coefficients of $h(x)$ may then be read off from the digits of $h(10^7)$, since the choice of evaluation point ensures that the coefficients do not overlap:
 \[ h(x) = 490590x^6 + 964034x^5 + 1043046x^4 + 1082839x^3 + 788467x^2 + 418982x + 151522. \]
Of course, on real hardware, one evaluates at $2^N$ rather than $10^N$, since then the packing and unpacking phases are accomplished efficiently (in linear time) by various bit shifting and masking operations.

The main advantage of this algorithm for multiplication in $\ZZ[x]$ is that it places the burden of computation on existing highly optimised software libraries for multiprecision integer arithmetic, such as the GMP library \cite{gmp}. This point of view is discussed further in \cite{fateman}. However, the algorithm also has a disadvantage: it introduces unwanted zero-padding. In the example above, if one computes the integer product by hand using the classical algorithm, about \emph{three-quarters} of the digit-by-digit products involve multiplying by zero, because about half of the input digits are zero. Clearly it is desirable to somehow skip these redundant multiplications.

In this paper we give several new algorithms that mitigate this inefficiency, without sacrificing the advantage mentioned above. The basic idea is to evaluate not just at a single point, but at several points; instead of reducing to a single multiplication, we reduce to several smaller multiplications. The evaluation points are chosen carefully so that the packing and unpacking phases are still highly efficient.

The benefit of this approach accrues as follows. Let $M(n)$ denote the time required to multiply $n$-bit integers. Suppose that the standard Kronecker scheme reduces a given polynomial multiplication problem in $\ZZ[x]$ to a multiplication of two $b$-bit integers in $\ZZ$. The cost is $M(b)$, plus $O(b)$ overhead associated with the packing and unpacking steps. One of our new algorithms (\S\ref{sec:integer-four}) reduces this problem instead to \emph{four} multiplications of size about $b/4$, so the cost becomes $4M(b/4) + O(b)$.

The relation between $M(b) + O(b)$ and $4M(b/4) + O(b)$ depends on the underlying integer multiplication algorithm, and on the implied constants in the $O(b)$ terms. If $b$ is very large, then one has available FFT-based methods for integer multiplication, and $M(b)$ behaves roughly like $b \log b$, leaving little difference between the two strategies. However, if $b$ is relatively small, then $M(b)$ will typically behave like $b^\alpha$ for some $\alpha > 1$; for example, GMP version 4.2.2 implements classical multiplication ($\alpha = 2$), Karatsuba multiplication ($\alpha \approx 1.58$), and Toom-Cook 3-way multiplication ($\alpha \approx 1.46$). In this situation we expect the new algorithm to win by a factor of about $4^{\alpha - 1}$. Under the classical multiplication regime, the theoretical gain is a factor of $4^{2-1} = 4$, which is equivalent to `skipping' all of the redundant multiplications by zero in the example given above. Finally, when $b$ is sufficiently small, we expect the $O(b)$ overhead to dominate, and the usual Kronecker substitution becomes faster, simply because it has a smaller constant in the $O(b)$ term. For the implementation discussed in \S\ref{sec:demo}, we find that the $O(b)$ term already interferes in the region where $\alpha$ is somewhat less than $2$, so we never quite achieve the factor of four speedup suggested by the above theoretical analysis.

Another interesting feature of the new algorithms is that they are trivially parallelisable, since they reduce the original problem to several independent multiplications. For example, using the algorithm described in \S\ref{sec:integer-four}, one can easily split a large multiplication problem in $\ZZ[x]$ into four threads, without needing to parallelise any of the internals of the integer multiplication routine.

The organisation of this paper is as follows. In \S\ref{sec:polynomial} we first consider the technically simplest case of reducing multiplication in $R[x, y]$ to multiplication in $R[x]$. The standard Kronecker substitution evaluates at $y = x^N$ for a suitably large $N$. We give three new algorithms. The first algorithm evaluates at $y = x^{N'}$ and $y = x^{-N'}$, where $N'$ is about half the size of $N$. The second algorithm evaluates at $y = x^{N'}$ and $y = -x^{N'}$, but only works for rings in which the multiply-by-two map is injective. The third algorithm combines these two algorithms, evaluating at four points $y = x^{N''}$, $x^{-N''}$, $-x^{N''}$ and $-x^{-N''}$, where $N''$ is about $N/4$. In \S\ref{sec:integer} we adapt these algorithms to the case of the substitution from $\ZZ[x]$ to $\ZZ$. The phenomenon of carries in integer arithmetic (the archimedean property of $\ZZ$) makes this slightly more complicated than the bivariate case. Finally in \S\ref{sec:demo} we give some examples of timings of an implementation of the algorithms for the specific problem of multiplication in $(\ZZ/n\ZZ)[x]$. We observe that the `four-point' variant is almost twice as fast as the standard Kronecker substitution, over a large range of problem sizes.

\section{The polynomial case}
\label{sec:polynomial}

Let $R$ be a commutative ring with identity. We will regard a polynomial $p \in R[x]$ as a vector of coefficients of a certain known length $\ell$ (i.e.~ the coefficient of $x^{\ell-1}$ is permitted to be zero). Similarly a bivariate polynomial $p \in R[x, y]$ will be treated as a rectangular array of coefficients, with a certain length $\ell_x$ with respect to $x$ and a certain length $\ell_y$ with respect to $y$. We write such a $p$ as $p = \sum_{i=0}^{\ell_y(p) - 1} p_i(x) y^i$, where each $p_i \in R[x]$ has length $\ell(p_i) = \ell_x(p)$.

Throughout this section we fix two polynomials $f, g \in R[x, y]$; we are interested in computing their product $h = fg$. For simplicity we assume that $\ell_x(f) = \ell_x(g)$ and that $\ell_y(f) = \ell_y(g)$. We denote these by $L_x$ and $L_y$ respectively, and assume that $L_x \geq 1$ and $L_y \geq 1$. We then have $\ell_x(h) = 2L_x - 1$ and $\ell_y(h) = 2L_y - 1$. It is not difficult to adapt all of the algorithms below to the case where $f$ and $g$ have different lengths.

\subsection{The standard Kronecker substitution}
\label{sec:polynomial-std}

Let $N = 2 L_x - 1$. We evaluate at $y = x^N$, that is, we compute
 \[ f(x, x^N) = \sum_{i=0}^{L_y - 1} f_i(x) x^{iN} \]
as an element of $R[x]$. Since $N \geq L_x$, this evaluation step consists simply of writing down the coefficients of $f_0, f_1, \ldots, f_{L_y - 1}$, with $L_x - 1$ zeroes between $f_i$ and $f_{i+1}$. The polynomial $f(x, x^N)$ has length $(L_y - 1)N + L_x = 2 L_x L_y - L_x - L_y + 1$. We evaluate similarly for $g$, and then multiply in $R[x]$ to obtain $h(x, x^N) = f(x, x^N) \cdot g(x, x^N)$. Note that
 \[ h(x, x^N) = \sum_{i=0}^{2L_y - 2} h_i(x) x^{iN}, \]
and $\ell(h_i) = 2L_x - 1 = N$ for each $i$. Therefore the coefficients of $h_i$ do not overlap those of $h_{i+1}$ in $h(x, x^N)$ for any $i$, so it is easy to read off the coefficients of $h(x, y)$. We obtain:

\begin{prop}
\label{prop:polynomial-std-ks}
The standard Kronecker substitution reduces the problem of computing $h = fg$ to multiplying two polynomials of length $2 L_x L_y - L_x - L_y + 1$ in $R[x]$.
\end{prop}

\subsection{Reciprocal evaluation points}
\label{sec:polynomial-recip}

In our first variant, we evaluate at $y = x^N$ and $x^{-N}$, where $N = L_x$. We obtain
\begin{align*}
  f(x, x^N) & = \sum_{i=0}^{L_y - 1} f_i(x) x^{iN}, \\
  x^{N(L_y - 1)} f(x, x^{-N}) & = \sum_{i=0}^{L_y - 1} f_i(x) x^{(L_y-1-i)N} = \sum_{i=0}^{L_y - 1} f_{L_y-1-i}(x) x^{iN}.
\end{align*}
(The normalising factor $x^{N(L_y - 1)}$ ensures that we have an element of $R[x]$ rather than of $R[x, x^{-1}]$.)

Since $N$ is exactly $L_x$, these two evaluations are obtained by concatenating the coefficients $f_0, f_1, \ldots, f_{L_y - 1}$ directly, with no zero-padding in between; they differ only in the order of concatenation.

We perform similar evaluations for $g$, and then compute the products
\begin{align*}
  h(x, x^N) & = f(x, x^N) \cdot g(x, x^N), \\
 x^{2N(L_y - 1)} h(x, x^{-N}) & = x^{N(L_y - 1)} f(x, x^{-N}) \cdot x^{N(L_y - 1)} g(x, x^{-N}).
\end{align*}
These are both multiplications of polynomials in $R[x]$ of length $N L_y = L_x L_y$.

We must now show how to recover the coefficients of $h(x, y)$ from knowledge of the two polynomials
\begin{align*}
  h(x, x^N) & = \sum_{i=0}^{2L_y - 2} h_i(x) x^{iN}, \\
 x^{2N(L_y - 1)} h(x, x^{-N}) & = \sum_{i=0}^{2L_y - 2} h_{2L_y-2-i}(x) x^{iN}.
\end{align*}
Note that each $h_i$ has length $2L_x - 1 = 2N - 1$, so the coefficients of $h_i$ generally \emph{do} overlap those of $h_{i+1}$, in both sums. However, there are two exceptions:
\begin{itemize}
\item The lowest $N$ terms of $h(x, x^N)$ are precisely the lowest $N$ terms of $h_0$.
\item The highest $N - 1$ terms of $x^{2N(L_y - 1)} h(x, x^{-N})$ are precisely the highest $N - 1$ terms of $h_0$.
\end{itemize}
We therefore completely recover $h_0$ by gluing together these two halves. Then we subtract $h_0$ from the appropriate position in both sums, revealing the two halves of $h_1$. We repeat this procedure for $h_1, h_2, \ldots, h_{2L_y - 2}$ to completely determine $h(x, y)$.

We obtain the following:
\begin{prop}
\label{prop:polynomial-recip-ks}
The `reciprocal' Kronecker substitution reduces the problem of computing $h = fg$ to two multiplications of polynomials of length $L_x L_y$ in $R[x]$, plus $O(L_x L_y)$ subtractions in $R$.
\end{prop}

\subsection{Negated evaluation points}
\label{sec:polynomial-negate}

The next algorithm only works for rings in which the multiply-by-two map is injective, so for example the algorithm does not work over a field of characteristic two.

As in \S\ref{sec:polynomial-recip}, put $N = L_x$. Evaluate at $y = x^N$ and $-x^N$, obtaining
 \[ f(x, x^N) = \sum_{i=0}^{L_y - 1} f_i(x) x^{iN}, \qquad  f(x, -x^N) = \sum_{i=0}^{L_y - 1} (-1)^i f_i(x) x^{iN}. \]
The first value is obtained by simply concatenating the coefficients of $f_0, f_1, \ldots, f_{L_y-1}$ without any intervening zero-padding. The second value is obtained in the same way, but with the sign alternating on each chunk.

Perform a similar evaluation on $g$, and then multiply to obtain $h(x, \pm x^N) = f(x, \pm x^N) \cdot g(x, \pm x^N)$. As in the `reciprocal' algorithm, these are both multiplications of polynomials of length $L_x L_y$.

Now decompose $h$ into even and odd parts $h^{(0)}$ and $h^{(1)}$ as
  \[ h(x, y) = h^{(0)}(x, y^2) + y h^{(1)}(x, y^2), \]
where $\ell_x(h^{(0)}) = \ell_x(h^{(1)}) = 2L_x - 1$, $\ell_y(h^{(0)}) = L_y$ and $\ell_y(h^{(1)}) = L_y - 1$. That is,
 \[  h^{(0)}(x, y) = \sum_{i=0}^{L_y-1} h_{2i}(x) y^i, \qquad h^{(1)}(x, y) = \sum_{i=0}^{L_y-2} h_{2i+1}(x) y^i. \]
We find that $h(x, \pm x^N) = h^{(0)}(x, x^{2N}) \pm x^N h^{(1)}(x, x^{2N})$, and inverting the system we obtain
\begin{align*}
  h^{(0)}(x, x^{2N}) & = \frac{h(x, x^N) + h(x, -x^N)}{2}, \\
  h^{(1)}(x, x^{2N}) & = \frac{h(x, x^N) - h(x, -x^N)}{2x^N}.
\end{align*}
Since $\ell_x(h_i) = 2L_x - 1 \leq 2N$, we are able to read off the coefficients of $h_{2i}$ from $h^{(0)}(x, x^{2N})$, and those of $h_{2i+1}$ from $h^{(1)}(x, x^{2N})$. Therefore we have:
\begin{prop}
\label{prop:polynomial-negate-ks}
The `negated' Kronecker substitution reduces the problem of computing $h = fg$ to two multiplications of polynomials of length $L_x L_y$ in $R[x]$, plus $O(L_x L_y)$ additions/subtractions in $R$ and $O(L_x L_y)$ divisions by $2$ in $R$.
\end{prop}

\subsection{Four evaluation points}
\label{sec:polynomial-four}

We continue to assume that doubling is injective in $R$. The final algorithm we present takes advantage of the fact that the key ideas of the `reciprocal' and `negated' variants are essentially orthogonal, and may be combined.

Let $N = \ceil{L_x/2}$. We evaluate at $y = x^N$, $-x^N$, $x^{-N}$ and $-x^{-N}$:
\begin{align*}
  f(x, x^N) & = \sum_{i=0}^{L_y - 1} f_i(x) x^{iN}, \\
  f(x, -x^N) & = \sum_{i=0}^{L_y - 1} (-1)^i f_i(x) x^{iN}, \\
  x^{N(L_y - 1)} f(x, x^{-N}) & = \sum_{i=0}^{L_y - 1} f_{L_y-1-i}(x) x^{iN}, \\
  x^{N(L_y - 1)} f(x, -x^{-N}) & = \sum_{i=0}^{L_y - 1} (-1)^i f_{L_y-1-i}(x) x^{iN}.
\end{align*}
A new phenomenon arises here: the coefficients $f_i$ overlap even in the \emph{evaluation} phase, in all four of the above sums, so some additions and subtractions in $R$ are required to compute them. Each of the four polynomials above has length $N(L_y - 1) + L_x$.

We evaluate similarly for $g$, and then multiply to obtain
\begin{align*}
  h(x, \pm x^N) & = f(x, \pm x^N) \cdot g(x, \pm x^N), \\
  x^{2N(L_y - 1)} h(x, \pm x^{-N}) & = x^{N(L_y - 1)} f(x, \pm x^{-N}) \cdot x^{N(L_y - 1)} g(x, \pm x^{-N}).
\end{align*}

As in \S\ref{sec:polynomial-negate}, we decompose $h(x, y)$ into odd and even parts,
  \[ h(x, y) = h^{(0)}(x, y^2) + y h^{(1)}(x, y^2), \]
and we find that
\begin{align*}
  h^{(0)}(x, x^{2N}) & = \frac{h(x, x^N) + h(x, -x^N)}{2}, \\
  h^{(1)}(x, x^{2N}) & = \frac{h(x, x^N) - h(x, -x^N)}{2x^N}, \\
  x^{2N(L_y - 1)} h^{(0)}(x, x^{-2N}) & = \frac{x^{2N(L_y - 1)} h(x, x^{-N}) + x^{2N(L_y - 1)} h(x, -x^{-N})}{2}, \\
  x^{2N(L_y - 2)} h^{(1)}(x, x^{-2N}) & = \frac{x^{2N(L_y - 1)} h(x, x^{-N}) - x^{2N(L_y - 1)} h(x, -x^{-N})}{2x^N}.
\end{align*}
(Note that $\ell_y(h^{(1)}) = L_y - 1$, whence the normalising factor $x^{2N(L_y-2)}$.)

Now consider $h^{(0)}(x, x^{2N})$ and $x^{2N(L_y - 1)} h^{(0)}(x, x^{-2N})$. Since $\ell(h_i) = 2L_x - 1 \leq 4N$, we may use the same strategy as in \S\ref{sec:polynomial-recip}, first reading off the low $2N$ coefficients of $h_0$ from $h^{(0)}(x, x^{2N})$ and the high $2N - 1$ coefficients of $h_0$ from $x^{2N(L_y - 1)} h^{(0)}(x, x^{-2N})$. Iterating, we obtain $h_2, h_4, \ldots, h_{2L_y-2}$. Applying the same procedure to $h^{(1)}(x, x^{2N})$ and $x^{2N(L_y - 2)} h^{(1)}(x, x^{-2N})$, we obtain $h_1, h_3, \ldots, h_{2L_y-3}$.

Finally we have:
\begin{prop}
\label{prop:polynomial-four-ks}
The `four-point' Kronecker substitution reduces the problem of computing $h = fg$ to four multiplications of polynomials of length $\ceil{L_x/2}(L_y - 1) + L_x$ in $R[x]$, plus $O(L_x L_y)$ additions/subtractions in $R$ and $O(L_x L_y)$ divisions by $2$ in $R$.
\end{prop}

\subsection{Complex evaluation points} We mention here another variant, which does not at present appear to be competitive with the algorithms above, but may suggest further avenues for research.

Put $M = \ceil{L_x/2}$, and let $S = R[i]$, where $i$ is a primitive fourth root of unity. We evaluate at $y = x^M$, $-x^M$ and $i x^M$. The value $f(x, i x^M) \in S[x]$ has `real' and `imaginary' parts, both lying in $R[x]$, so we have four polynomials in $R[x]$ altogether. We compute the pointwise products,
\begin{align*}
  h(x, x^M) & = f(x, x^M) \cdot g(x, x^M), \\
  h(x, -x^M) & = f(x, -x^M) \cdot g(x, -x^M), \\
  h(x, ix^M) & = f(x, ix^M) \cdot g(x, ix^M).
\end{align*}
The first two multiplications are ordinary multiplications in $R[x]$; the last one is a `complex' multiplication, and so requires three multiplications in $R[x]$. One checks that by taking appropriate linear combinations of these three products, the coefficients of $h$ may be reconstructed. Unfortunately, one has committed five multiplications instead of four, so the algorithm appears to be inferior to the algorithm of the previous section.

Further variants are possible. For example, take $S = R[\omega]$, where $\omega$ is a primitive cube root of unity, and let $M = \ceil{L_x/3}$. One may evaluate at $y = \pm x^M$ and $y = \pm \omega x^M$, reducing the problem to eight multiplications in $R[x]$ that are one-sixth the size of the multiplication generated by the standard Kronecker substitution.

\section{The integer case}
\label{sec:integer}

In this section we adapt the above algorithms to the case of the substitution from $\ZZ[x]$ to $\ZZ$. This is mostly straightforward; the main complication is the management of carries in the reconstruction phase in the analogue of the `reciprocal' algorithm.

Fix two polynomials $f, g \in \ZZ[x]$, with product $h$. We assume that they have the same length $L$, and put $e = \ceil{\log_2 L}$. We write $f = \sum_{i=0}^{L-1} f_i x^i$, and similarly for $g$. The \emph{length} of an integer $n$ is defined to be $1 + \floor{\log_2|n|}$ (the number of bits in the binary representation of $|n|$). We assume that $f$ and $g$ have \emph{non-negative} coefficients of length at most $b$ for some integer $b \geq 1$. It should be possible to handle the case of signed coefficients using essentially the same techniques, but we have not checked the details.

Note that the coefficients of the product $h$ have length at most $2b + e$. In fact, they satisfy the slightly stronger inequality
\begin{equation}
\label{eq:h-inequality}
 0 \leq h_i \leq L (2^b - 1)^2 \leq 2^e (2^{2b} - 2^{b+1} + 1),
\end{equation}
which we will need in \S\ref{sec:integer-recip} to control the propagation of carries.

We assume that integers of length $n$ may be added and subtracted in time $O(n)$, and that we may divide by a power of two in time $O(n)$. We also assume that we can `pack' and `unpack' binary strings in linear time. More precisely, given a list of integers $a_0, \ldots, a_{k-1}$ satisfying $0 \leq a_i < 2^c$ for some integer $c$, we require that we can construct the sum $\sum_{i=0}^{k-1} a_i 2^{ic}$ in time $O(kc)$, and given this packed representation, we require that we can reconstruct the sequence of $a_i$ in time $O(kc)$.

\subsection{The standard Kronecker substitution}
\label{sec:integer-std}

Let $N = 2b + e$, so that the coefficients of $h$ have length at most $N$. Evaluate at $x = 2^N$ to obtain $f(2^N)$ and $g(2^N)$, multiply to obtain $h(2^N) = f(2^N) g(2^N)$, and unpack $h(2^N)$ to obtain the $h_i$.

\begin{prop}
The standard Kronecker substitution reduces the problem of computing $h = fg$ to multiplying two integers of length $(2b+e)(L-1) + b$, plus packing/unpacking overhead of $O((2b+e)L)$.
\end{prop}

\subsection{Reciprocal evaluation points}
\label{sec:integer-recip}

Let $N = \ceil{(2b+e)/2} = b + \ceil{e/2}$. We evaluate at $x = 2^N$ and $2^{-N}$:
\begin{align*}
 f(2^N) & = \sum_{i=0}^{L-1} f_i 2^{iN}, \\
 2^{N(L-1)} f(2^{-N}) & = \sum_{i=0}^{L-1} f_i 2^{(L-1-i)N} = \sum_{i=0}^{L-1} f_{L-1-i} 2^{iN}.
\end{align*}
Note that there are only $\ceil{e/2}$ bits of zero-padding between adjacent coefficients in the above sums. We evaluate similarly for $g$, and then compute the integer products
\begin{align*}
  h(2^N) & = f(2^N) \cdot g(2^N), \\
  2^{2N(L-1)} h(2^{-N}) & = 2^{N(L-1)} f(2^{-N}) \cdot 2^{N(L-1)} g(2^{-N}).
\end{align*}

Now we must show how to recover the $h_i$ from the two sums
\begin{align}
 h(2^N) & = \sum_{i=0}^{2L-2} h_i 2^{iN}, \label{eq:overlap-1} \\
 2^{2N(L-1)} h(2^{-N}) & = \sum_{i=0}^{2L-2} h_{2L-2-i} 2^{iN}. \label{eq:overlap-2}
\end{align}
The $h_i$ overlap in both sums, so to retrieve them we must use a similar strategy to that described in \S\ref{sec:polynomial-recip} for the polynomial case. There are added complications due to the presence of carries, which we resolve as follows.

First we write both sums in base $2^N$. Note that $f(2^N)$ and $g(2^N)$ have length at most $LN$, so $h(2^N)$ has length at most $2LN$. Therefore we may write
\begin{equation}
\label{eq:digits-1}
 h(2^N) = \sum_{i=0}^{2L-1} u_i 2^{iN}
\end{equation}
where each digit $u_i$ satisfies $0 \leq u_i < 2^N$. Similarly $2^{2N(L-1)} h(2^{-N})$ has length at most $2NL$, so we may write
\begin{equation}
\label{eq:digits-2}
 2^{2N(L-1)} h(2^{-N}) = \sum_{i=0}^{2L-1} w_{2L-1-i} 2^{iN}
\end{equation}
where $0 \leq w_i < 2^N$. Decompose each $h_i$ into two digits as
 \[ h_i = \alpha_i + 2^N \beta_i, \qquad 0 \leq i \leq 2L-2, \]
where
\begin{align}
 0 & \leq \alpha_i < 2^N, \notag \\
 0 & \leq \beta_i < 2^N - 1. \label{eq:beta-range}
\end{align}
The latter inequality is equivalent to saying that $h_i < 2^N(2^N - 1)$, which follows from \eqref{eq:h-inequality} since
\begin{align*}
 h_i & \leq 2^{2b+e} - 2^{b+e+1} + 2^e \\
     & = 2^{2b+e} - 2^{b+e} - 2^e(2^b - 1) \\
     & < 2^{2b+e} - 2^{b+e} \\
     & \leq 2^{2N} - 2^N.
\end{align*}

The various quantities we have introduced satisfy the following relations. From \eqref{eq:overlap-1} and \eqref{eq:digits-1} we have $\alpha_0 = u_0$, and
\begin{equation}
\label{eq:carry-1}
 \beta_i + \alpha_{i+1} + \delta_i = u_{i+1} + 2^N \delta_{i+1},  \qquad 0 \leq i \leq 2L-2,
\end{equation}
where $\delta_0 = \alpha_{2L-1} = 0$ and where $\delta_{i+1} \in \{0, 1\}$ is the carry generated by the addition $\beta_i + \alpha_{i+1} + \delta_i$. Similarly, from \eqref{eq:overlap-2} and \eqref{eq:digits-2} we have $\alpha_{2L-2} = w_{2L-1}$, and
\begin{equation}
\label{eq:carry-2}
 \alpha_i + \beta_{i+1} + \ep_{i+1} = w_{i+1} + 2^N \ep_i, \qquad -1 \leq i \leq 2L-3,
\end{equation}
where $\ep_{2L-2} = \alpha_{-1} = 0$ and $\ep_i \in \{0, 1\}$ is the carry generated by the addition $\alpha_i + \beta_{i+1} + \ep_{i+1}$. Note that $\ep_{-1} = 0$ since we know that $2^{2N(L-1)} h(2^{-N})$ fits into $2L$ digits.

Given the $u_i$ and $w_i$, we solve these equations for $\alpha_i$ and $\beta_i$ (and incidentally for $\delta_i$ and $\ep_i$) by the following iterative procedure. Start with $\alpha_0 = u_0$ and $\alpha_{-1} = \delta_0 = \ep_{-1} = 0$. Now let $0 \leq j \leq 2L - 3$, and suppose that $\alpha_{j-1}$, $\alpha_j$, $\delta_j$ and $\ep_{j-1}$ have been computed. From \eqref{eq:carry-2} we have
 \[ \beta_{j+1} + \ep_{j+1} = (w_{j+1} - \alpha_j) + 2^N \ep_j. \]
By \eqref{eq:beta-range}, the left hand side is less than $2^N$. Therefore, $\ep_j = 1$ if $\alpha_j > w_{j+1}$, and $\ep_j = 0$ otherwise. Taking \eqref{eq:carry-2} modulo $2^N$ for $i = j - 1$, we deduce the value of $\beta_j$:
 \[ \beta_j = w_j - \alpha_{j-1} - \ep_j \pmod{2^N}. \]
From \eqref{eq:carry-1} modulo $2^N$ we obtain $\alpha_{j+1}$,
 \[ \alpha_{j+1} = u_{j+1} - \beta_j - \delta_j  \pmod{2^N}, \]
and then $\delta_{j+1}$ is obtained directly from \eqref{eq:carry-1}. At this stage we have found $\alpha_j$, $\alpha_{j+1}$, $\delta_{j+1}$ and $\ep_j$, and so we may repeat the process, until we have determined $\alpha_0, \ldots, \alpha_{2L-2}$ and $\beta_0, \ldots, \beta_{2L-3}$. Finally we obtain $\beta_{2L-2}$ from \eqref{eq:carry-2} with $i = 2L-3$:
 \[  \beta_{2L-2} = w_{2L-2} - \alpha_{2L-3} - \ep_{2L-2} \pmod{2^N}. \]
The $h_i$ are then reconstructed as $h_i = \alpha_i + 2^N \beta_i$.

\begin{prop}
\label{prop:integer-recip-ks}
The `reciprocal' Kronecker substitution reduces the problem of computing $h = fg$ to two multiplications of integers of length $(b+\ceil{e/2})(L-1) + b$, plus packing/unpacking overhead of $O((2b+e)L)$.
\end{prop}

\begin{example}
We illustrate (in base ten) using the example from \S\ref{sec:intro}. We have $N = 4$, so the evaluation points are $x = 10^4$ and $x = 10^{-4}$. Let
\begin{align*}
  f(x) & = 621x^3 + 887x^2 + 610x + 274, \\
  g(x) & = 790x^3 + 424x^2 + 298x + 553, \\
  h(x) & = f(x) g(x).
\end{align*}
We have
\begin{align*}
  f(10^4) & = 621|0887|0610|0274, &  10^{12} f(10^{-4}) & = 274|0610|0887|0621, \\
  g(10^4) & = 790|0424|0298|0553, &  10^{12} g(10^{-4}) & = 553|0298|0424|0790,
\end{align*}
with the vertical bars showing boundaries between base-$10^4$ digits.
The pointwise products are
\begin{align*}
  h(10^4) & = f(10^4) g(10^4) \\
          & = 49|0686|4138|3154|2917|8508|8997|1522, \\
  10^{24} h(10^{-4}) & = 10^{12} f(10^{-4}) \cdot 10^{12} g(10^{-4}) \\
          & = 15|1563|9060|8575|2943|3142|4083|0590.
\end{align*}
The low digit $u_0 = 1522$ of $h(10^4)$ is the bottom half $\alpha_0$ of $h_0$. Comparing $1522$ to $w_1 = 1563$, we see that there was no carry, i.e. $\ep_0 = 0$. Thus the top half $\beta_0$ is simply $w_0 = 15$, so $h_0 = 151522$. We will not follow through the rest of the algorithm in detail, but we can see that by subtracting $151522$ from the appropriate positions in both values, we reveal the top and bottom halves of $h_1 = 418982$:
\begin{align*}
  h(10^4) - 151522 & = 49|0686|4138|3154|2917|8508|8982|0000, \\
  10^{24} h(10^{-4}) - 10^{24} 151522 & = 00|0041|9060|8575|2943|3142|4083|0590.
\end{align*}

\end{example}

\subsection{Negated evaluation points}
\label{sec:integer-negate}

Put $N = b + \ceil{e/2}$ and evaluate at $2^N$ and $-2^N$. In $f(2^N)$, there are $\ceil{e/2}$ bits of zero-padding between adjacent coefficients; in $f(-2^N)$ the padding alternates between zero-padding and `one-padding'. If one only has available a `packing routine' for non-negative inputs, $f(-2^N)$ may be determined by first computing $f^{(0)}(2^{2N})$ and $2^N f^{(1)}(2^{2N})$ separately, where $f^{(0)}$ and $f^{(1)}$ are the even and odd parts of $f$, and then taking their difference. Note that $f(-2^N)$ may be negative, if the leading monomial of $f$ has odd exponent.

\begin{prop}
\label{prop:integer-negate-ks}
The `negated' Kronecker substitution reduces the problem of computing $h = fg$ to two multiplications of integers of length $(b + \ceil{e/2})(L-1) + b$, plus packing/unpacking overhead of $O((2b+e)L)$.
\end{prop}

\begin{example}
We illustrate with our running example. The evaluation points are $x = 10^4$ and $x = -10^4$:
\begin{align*}
  f(10^4) & = 887|00000274 + 621|00000610|0000 = 621088706100274, \\
  f(-10^4) & = 887|00000274 - 621|00000610|0000 = -620911306099726, \\
  g(10^4) & = 424|00000553 + 790|00000298|0000 = 790042402980553, \\
  g(-10^4) & = 424|00000553 - 790|00000298|0000 = -789957602979447.
\end{align*}
The pointwise products are
\begin{align*}
  h(10^4) & = f(10^4) g(10^4) = 490686413831542917850889971522, \\
  h(-10^4) & = f(-10^4) g(-10^4) = 490493607029377239842510331522.
\end{align*}
The even and odd coefficients of $h$ are then read off from
\begin{align*}
  h^{(0)}(10^8) & = (h(10^4) + h(-10^4))/2 = 490590|01043046|00788467|00151522, \\
  10^4 h^{(1)}(10^8) & = (h(10^4) - h(-10^4))/2 = 964034|01082839|00418982|0000.
\end{align*}
\end{example}

\subsection{Four evaluation points}
\label{sec:integer-four}

We take $N = \ceil{(2b+e)/4}$, and evaluate at $y = 2^N$, $-2^N$, $2^{-N}$ and $-2^{-N}$. The structure of the algorithm is the same as that of \S\ref{sec:polynomial-four}, and we omit the details. The reconstruction algorithm of \S\ref{sec:integer-recip} must be used twice: first on $h^{(0)}(2^{2N})$ and $2^{2N(L - 1)} h^{(0)}(2^{-2N})$ to recover the even-index coefficients of $h$, and then on $h^{(1)}(2^{2N})$ and $2^{2N(L - 2)} h^{(1)}(2^{-2N})$ to recover the odd-index coefficients.

\begin{prop}
\label{prop:integer-four-ks}
The `four-point' Kronecker substitution reduces the problem of computing $h = fg$ to four multiplications of integers of length $\ceil{(2b+e)/4}(L-1) + b$, plus packing/unpacking overhead of $O((2b+e)L)$.
\end{prop}

\begin{example}
Continuing with the running example, we put $N = 2$. For $f$ we have
\begin{align*}
  f(10^2) & = 887|0274 + 621|0610|00 = 629931274, \\
  f(-10^2) & = 887|0274 - 621|0610|00 = -612190726, \\
  10^6 f(10^{-2}) & = 274|0887|00 + 610|0621 = 280189321, \\
  10^6 f(-10^{-2}) & = 274|0887|00 - 610|0621 = 267988079,
\end{align*}
and for $g$ we have
\begin{align*}
  g(10^2) & = 424|0553 + 790|0298|00 = 794270353, \\
  g(-10^2) & = 424|0553 - 790|0298|00 = -785789247, \\
  10^6 g(10^{-2}) & = 553|0424|00 + 298|0790 = 556023190, \\
  10^6 g(-10^{-2}) & = 553|0424|00 - 298|0790 = 550061610.
\end{align*}
The pointwise products are
\begin{align*}
  h(10^2) & = 500335735365719722, \\
  h(-10^2) & = 481052889603923322, \\
  10^{12} h(10^{-2}) & = 155791760066353990, \\
  10^{12} h(-10^{-2}) & = 147409954195547190.
\end{align*}
Then we obtain
\begin{align*}
  h^{(0)}(10^4) & = (h(10^2) + h(-10^2))/2 & = 49|0694|3124|8482|1522, \\
  10^{12} h^{(0)}(10^{-4}) & = (10^{12} h(10^{-2}) + 10^{12} h(-10^{-2}))/2 & = 15|1600|8571|3095|0590,
\end{align*}
from which we recover the even-index coefficients of $h$, using the reconstruction algorithm from \S\ref{sec:integer-recip}. The odd-index coefficients are similarly found from
\begin{align*}
  10^2 h^{(1)}(10^4) & = (h(10^2) - h(-10^2))/2 & = 96|4142|2880|8982|00, \\
  10^{10} h^{(1)}(10^{-4}) & = (10^{12} h(10^{-2}) - 10^{12} h(-10^{-2}))/2 & = 41|9090|2935|4034|00.
\end{align*}
\end{example}

\section{Example timings}
\label{sec:demo}

The author implemented the algorithms in C for the case of multiplication in $(\ZZ/n\ZZ)[x]$, where $n$ fits into a single machine word. More precisely, the implementation first lifts the input polynomials to $\ZZ[x]$, multiplies in $\ZZ[x]$ using one of the algorithms of \S\ref{sec:integer}, and then reduces the result modulo $n$. The underlying integer arithmetic is performed by GMP's low-level `mpn' routines. The code is freely available under the GNU General Public License (GPL) from the author's web site, \texttt{http://math.harvard.edu/$\sim$dmharvey/}.

The timing data shown below were obtained on a 1.8GHz 64-bit AMD Opteron machine, kindly supplied by William Stein (funded by NSF grant No.~0555776). Both our code and GMP 4.2.1 were compiled using gcc 4.1.2, with the \texttt{-O2} optimisation flag. We also used Pierrick Gaudry's AMD patch for GMP, which improves the performance of GMP on the Opteron.

\begin{figure}
\begin{center}
\includegraphics[width=0.9\textwidth]{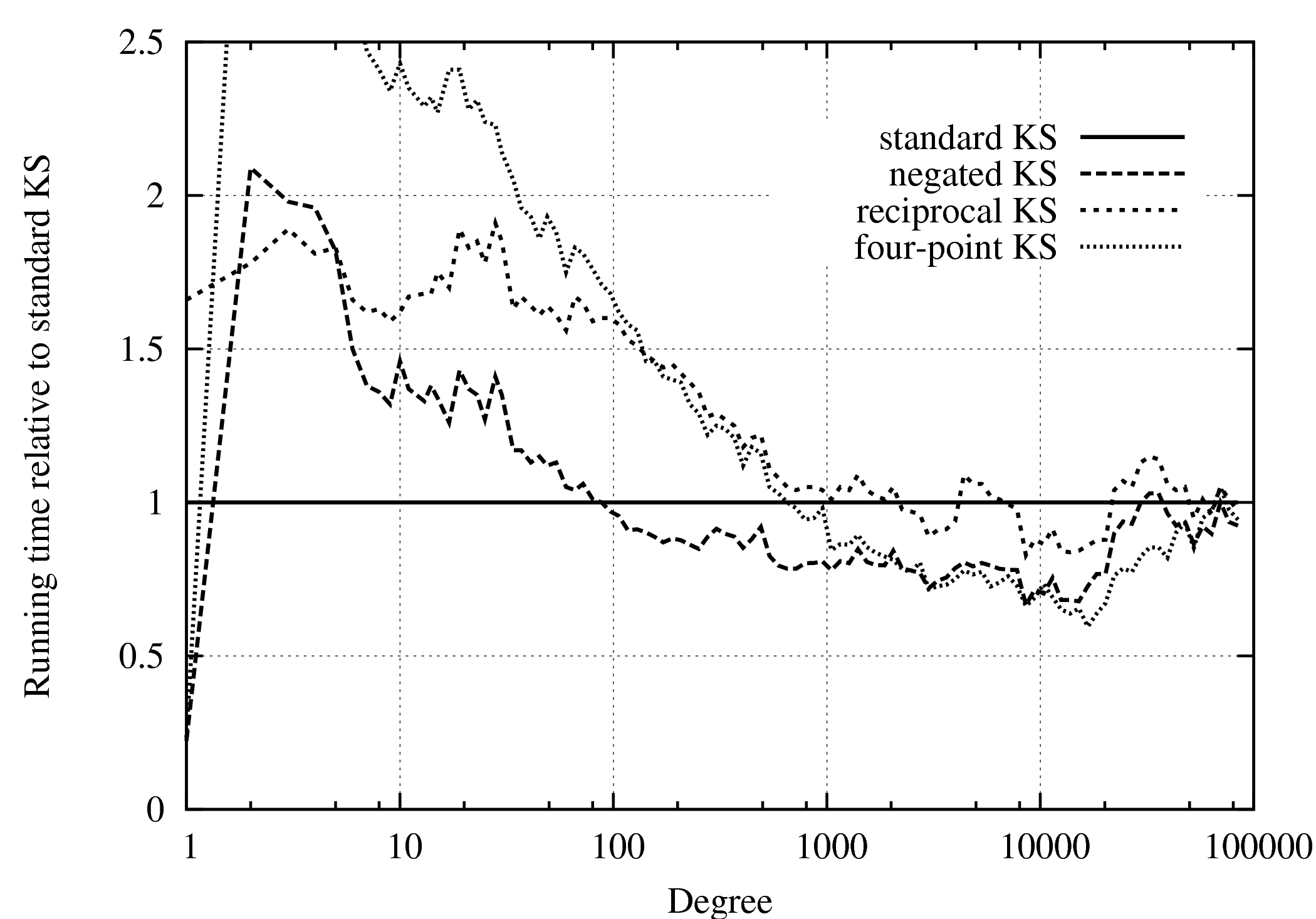}
\caption{Comparison of the four algorithms for a 4-bit modulus}
\label{fig:4-bit}
\end{center}
\end{figure}

\begin{figure}
\begin{center}
\includegraphics[width=0.9\textwidth]{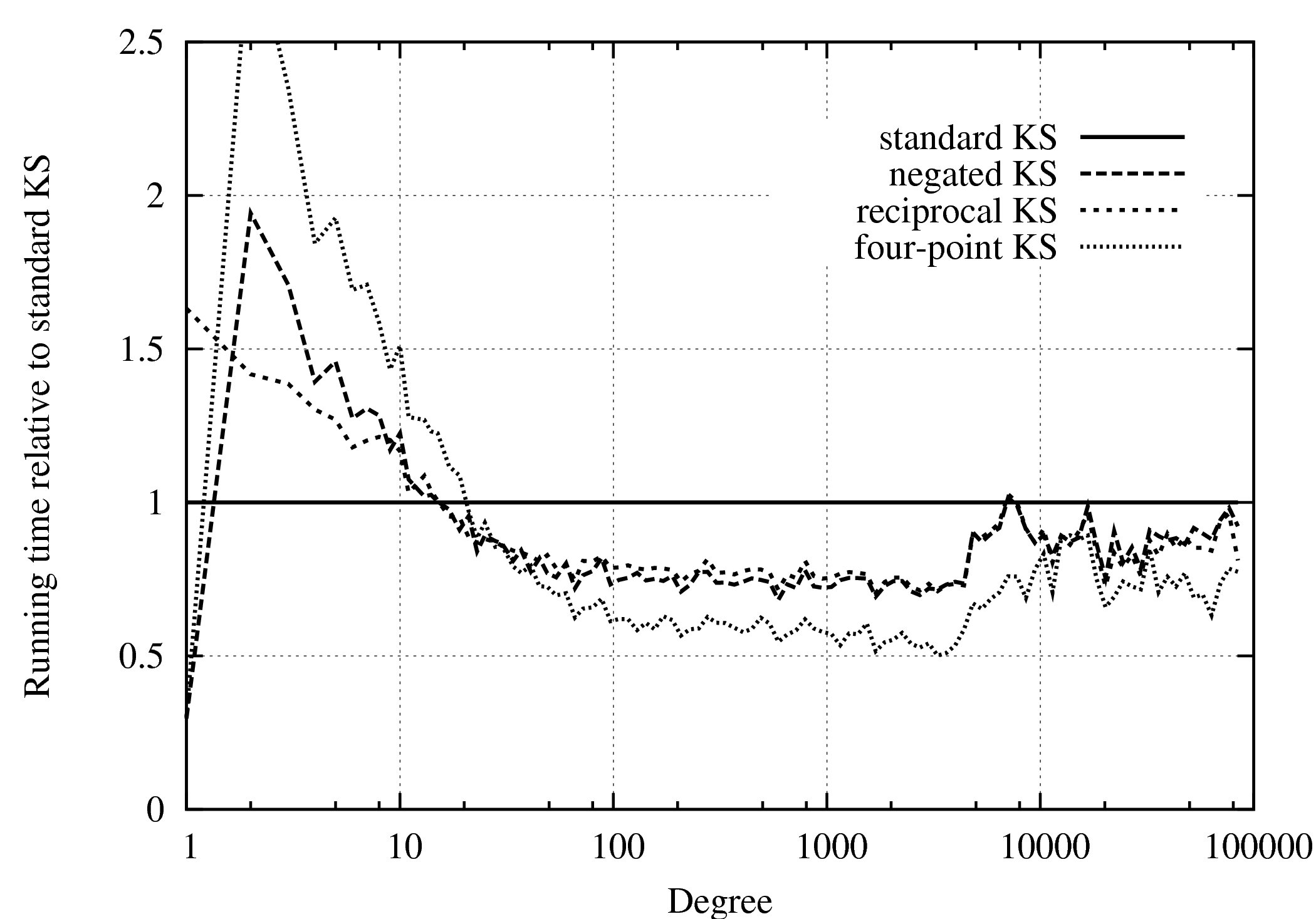}
\caption{Comparison of the four algorithms for a 48-bit modulus}
\label{fig:48-bit}
\end{center}
\end{figure}

Figure \ref{fig:4-bit} shows the relative performance of the three new algorithms (\S\ref{sec:integer-recip}, \S\ref{sec:integer-negate}, \S\ref{sec:integer-four}) compared to the standard Kronecker substitution (\S\ref{sec:integer-std}), where $n$ is a random 4-bit modulus. Figure \ref{fig:48-bit} is the same, but for a 48-bit modulus. We note several interesting features of the graphs:

\begin{itemize}
\item On both graphs, the four curves converge towards $1$ as the degree grows. This reflects the asymptotically quasilinear running time of the underlying integer multiplication routine, as discussed in \S\ref{sec:intro}.
\item The most impressive region is in Figure \ref{fig:48-bit}, between degrees roughly 100 and 5000. In this range, the four-point variant is almost twice as fast as the standard Kronecker substitution.
\item On both graphs, the negated variant has better performance than the reciprocal variant (although the difference is marginal in the 48-bit case). This is due to the added overhead of the complicated reconstruction algorithm of \S\ref{sec:integer-recip}.
\item The new algorithms gain more over the standard Kronecker substitution in the 48-bit modulus case than in the 4-bit modulus case. This occurs because the packing/unpacking overhead takes up a larger proportion of the total time in the 4-bit case.
\item For sufficiently small degree, the new algorithms are inferior to the standard Kronecker substitution, due to packing/unpacking overhead.
\end{itemize}

\begin{figure}
\begin{center}
\includegraphics[width=0.9\textwidth]{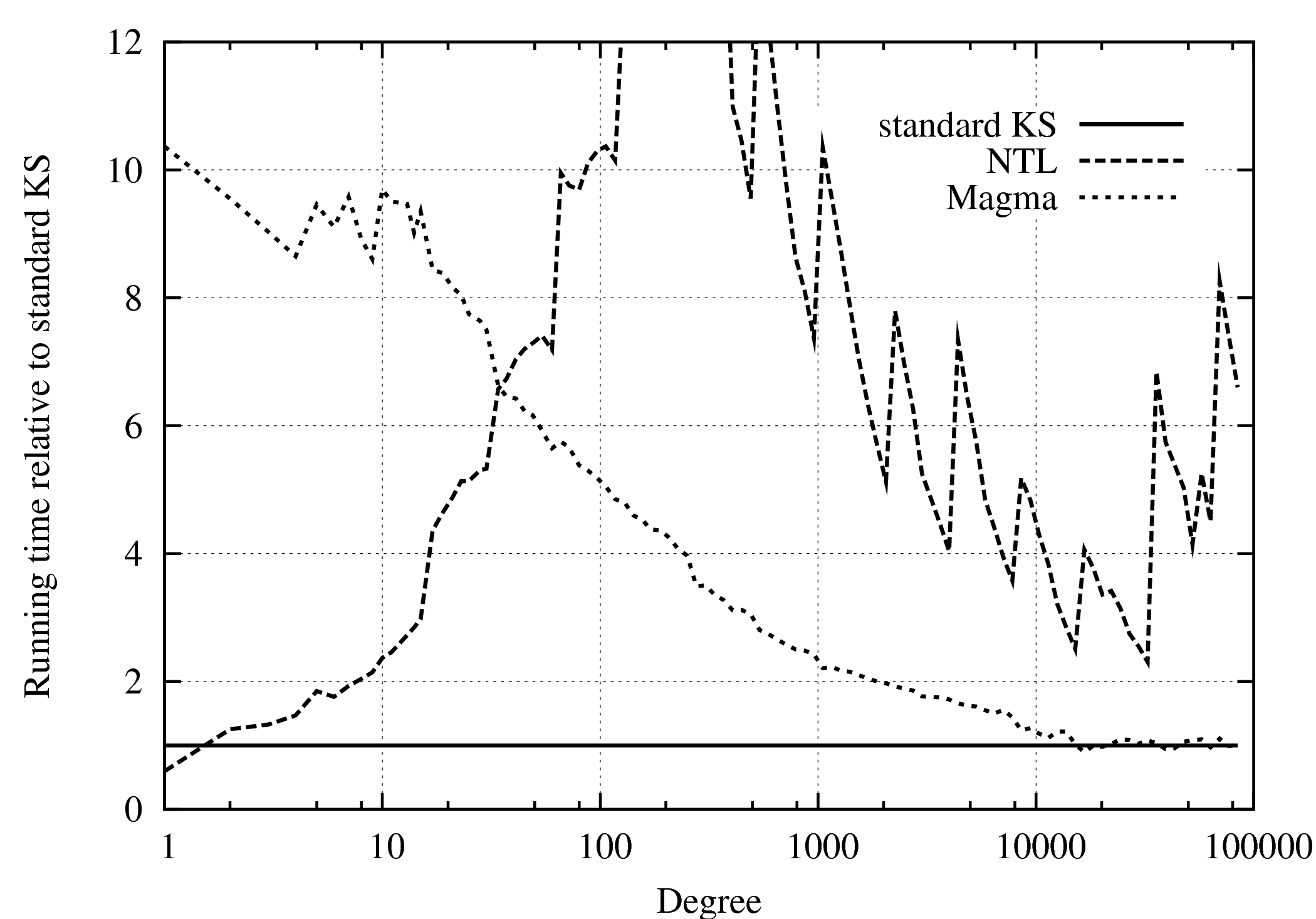}
\caption{Comparison with other systems for a 4-bit modulus}
\label{fig:4-bit-compare}
\end{center}
\end{figure}

\begin{figure}
\begin{center}
\includegraphics[width=0.9\textwidth]{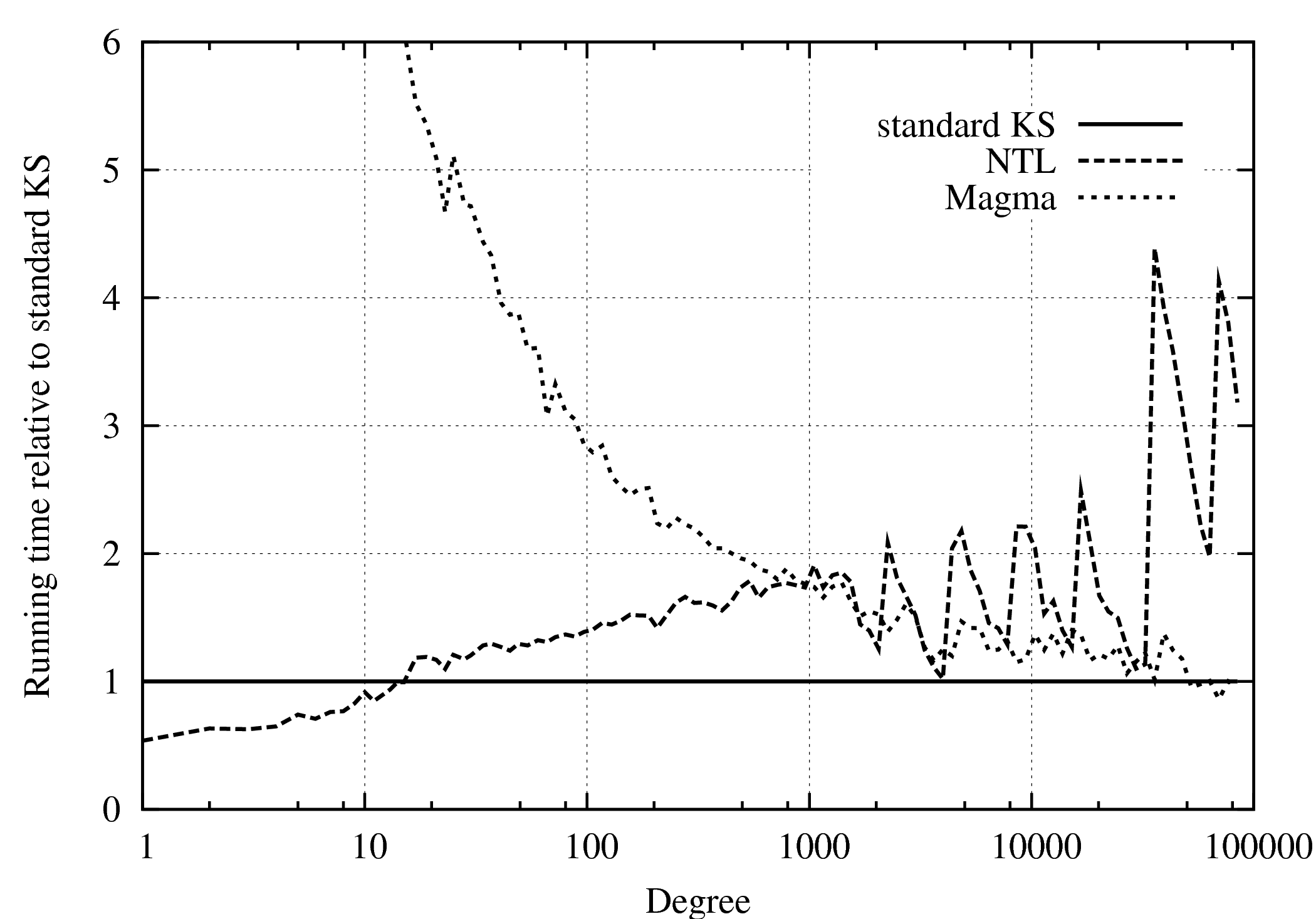}
\caption{Comparison with other systems for a 48-bit modulus}
\label{fig:48-bit-compare}
\end{center}
\end{figure}

For reference, we also compared the performance of our code on the same machine to two well-known systems, Magma (version 2.13-5) and NTL (version 5.4.1). The latter has specialised routines for arithmetic on polynomials with word-sized coefficients (the \texttt{zz\_pX} class). Figure \ref{fig:4-bit-compare} compares our implementation of the \emph{standard} Kronecker substitution against both Magma and NTL for a 4-bit modulus, and Figure \ref{fig:48-bit-compare} is for a 48-bit modulus.

\section*{Acknowledgements}

Many thanks to Paul Zimmermann, Andrew Sutherland and William Hart for their comments on this paper, and for stimulating conversations about these algorithms.

\bibliographystyle{amsplain}
\bibliography{kronecker}

\newpage

\end{document}